\documentstyle[aps,prl,multicol,epsfig]{revtex} 
\renewcommand{\narrowtext}{\begin{multicols}{2} \global\columnwidth20.5pc}
\renewcommand{\widetext}{\end{multicols} \global\columnwidth42.5pc}
\newcommand{\Lrule}{\vspace*{-0.2in}\noindent\vrule width3.5in height.2pt
  depth.2pt \vrule depth0em height1em}
\newcommand{\Rrule}{\vspace{-0.1in}\hfill\vrule depth1em height0pt \vrule
  width3.5in height.2pt depth.2pt\vspace*{-0.1in}}

\newcommand{\MeV}{{\rm MeV}}
\newcommand{\GeV}{{\rm GeV}}

\newcommand{\fm}{{\rm fm}}
\renewcommand{\Im}{{\rm Im}}

\begin{document}
\draft
%\wideabs{
\title{ $J/\psi$ and $\eta_{c}$ in the nuclear medium: QCD sum rule
 approach\thanks{work supported in part by BMBF, GSI, KOSEF and BSRI-98-2425}}
\author{Frank Klingl$^1$, Sungsik Kim$^3$, Su Houng
 Lee$^{1,2,3}$\footnote{Alexander von Humboldt fellow}, Philippe Morath$^1$ and Wolfram Weise$^1$}
\address{$^1$ Physik-Department, Technische Universit\"at M\"unchen, D-85747
 Garching, Germany  \\ $^2$ GSI, D-64291 Darmstadt, Germany \\
 $^3$ Department of Physics, Yonsei University, Seoul 120-749, Korea }
\maketitle
\begin{abstract}
We investigate the masses of the lowest $c \bar{c}$ states, the $J/\psi$ and
$\eta_c$, in nuclear matter using QCD sum rules. Up to dimension four, the
differences between the operator product expansions in vacuum and in medium
arise from the density-dependent change in the gluon condensate and from a new
contribution proportional to the nucleon expectation value of the twist-2 gluon
operator. Both terms together give an attractive shift of about 5-10 MeV to
the $J/\psi$ and $\eta_c$ masses in nuclear matter.
\end{abstract}
%}
\narrowtext

Investigating the behaviour of heavy quark systems in a nuclear medium is of
great interest, for several reasons. First, the ongoing discussion of $J/\psi$
suppression in ultrarelativisitic heavy-ion collisions as a possible
quark-gluon plasma signal requires detailed knowledge about the in-medium
interactions of the $J/\psi$ under ``normal'', non-plasma
conditions. Furthermore, as Brodsky et al \cite{Brodsky+90} pointed out,
multigluon exchange can lead to an attractive potential between a $c
\bar{c}$-meson and a nucleon, such that, for example, the $\eta_c$ could form
bound states even with light nuclei. In more recent calculations
the estimated charmonium binding energy in nuclear systems was
found to be of the order of 10 \MeV \cite{Wasson91,Luke+92,Brodsky+97,Teramond+98}.
 
In the present paper we study the in-medium behaviour of the $J/\psi$ and
$\eta_c$ using QCD sum rules\cite{Shifman+79}. The QCD sum rule approach
connects the spectral density of a given current correlation function via a
dispersion relation with the QCD operator product expansion (OPE). 
In-medium QCD sum rules have so far been applied only for light quark systems,
 in order to study possible shifts of the in-medium masses of 
nucleons \cite{DL91,Hat91,Tom91} and  vector mesons\cite{HL92}. 
Such calculations suffer from uncertainties, e.g. due to
assumptions about factorization of four-quark condensates which may not be
justified. As we shall see, in-medium QCD sum rules applied to heavy quark
systems are expected to be more reliable. Up to dimension four, the order to
which the vacuum sum rules for hadrons involving heavy quarks are commonly
expanded, all condensate parameters are quite well known and there are no
ambiguities in the OPE. We also find that uncertainties caused by possibly
large hadronic in-medium decay widths are much smaller than for light-quark
systems. 

Our starting point is the time ordered current-current correlation function of two heavy quark currents in nuclear matter,
\begin{equation}
\label{eq1}
\Pi (\omega, \vec{q}\,)=i\int d^4x \,e^{iq \cdot x}\langle |T[j(x)j(0)]|\rangle_{n.m.}.
\end{equation}
Here $q=( \omega, \vec{q})$, and $| \rangle_{n.m.}$ is the ground state of
nuclear matter which we take to be at rest. For the $J/\psi$ we take the vector current $j^{V}_{\mu}=\bar{c}\gamma_{\mu}c$ and 
for the $\eta_c$, we use the pseudoscalar current $j^P=i\bar{c}\gamma_{5}c$. 
In the region of large and positive $Q^2=\vec{q}^{\, 2}-\omega^2$ we can 
express the correlation function through an operator product expansion 
(short distance expansion)\cite{Wilson69} and 
write the left hand side of eq.(\ref{eq1}) as 
\begin{eqnarray}
\label{eq2}
\Pi (\omega, \vec{q}\,)=\sum_n C_{n}(\omega,\vec{q}) \, 
\langle O_{n} \rangle.
\end{eqnarray}
Here the $O_n$ are operators of (mass) dimension $n$, renormalized at 
a scale $\mu^2$, and $C_{n}$ are the perturbative Wilson 
coefficients.

At baryon densities $\rho_N$ for which the chemical potential is small compared
to the scale $\mu$ separating short and long distance phenomena, 
all density effects can be put into the $\rho_N$ dependence of the condensates
$\langle O_{2n} \rangle$, and we can use the perturbative Wilson coefficients calculated in the vacuum \cite{HL92,Tom95}.  
In heavy quark systems the expansion of quark operators in terms of inverse
powers of the large quark mass permits to express them entirely in terms of gluonic operators\cite{Shifman+79,Reinders+81,Reinders+85}.  In
the vacuum 
only the scalar gluon condensate $\langle \frac{\alpha_s}{\pi} G_{\mu \nu}
G^{\mu \nu} \rangle$
contributes up to dimension four.  In nuclear matter, an additional
contribution involving in-medium expectation values of the twist-2 tensorial
gluon operator $\langle \frac{\alpha_s}{\pi} G_{\alpha \sigma}G_\beta^\sigma
\rangle$ enters. We discuss this new term in some detail.

We will use  the 
linear, low-density approximation \cite{Cohen+92} for the in-medium
condensates:  
\begin{equation}
\label{eq3}
\langle O \rangle_{n.m.}=\langle O\rangle_{0}+
\frac{\rho_N}{2m_{N}}\langle N|O|N \rangle
\end{equation}
where $ \langle \rangle_0$ represents the vacuum expectation value, and 
the nucleon state (taken at rest in eq.~(\ref{eq3})) is  normalized as 
$\langle N(p')|N(p)\rangle=2p_{0}(2\pi)^3\delta^{3}(\vec{p}-\vec{p}\,')$. 
The in-medium changes of the condensates can then be related to the nucleon
expectation values of the corresponding operators. For the traceless and
symmetric gluonic twist-2 tensor operator we write 
\begin{equation}
\langle N(p)|\frac{\alpha_s}{\pi}G^{\alpha \sigma} 
G^\beta_{\,\, \,\sigma} | N(p) \rangle
=-(p^{\alpha}p^{\beta}-\frac{1}{4}g^{\alpha \beta}p^2)
\frac{\alpha_s}{\pi}A_{G}
\end{equation} 
where $m_{N}$ is the nucleon mass and $A_{G}$ is related to the following
moment of the gluon distribution function $G$:
\begin{equation}
A_{G}(\mu^2)=2\int_{0}^{1}dx \,x \, G(x,\mu^2).
\end{equation}
It represents twice the momentum fraction carried by gluons in the nucleon. We
take $A_{G}(8 m_c^2) \simeq 0.9$ \cite{Gluck+} at the scale $\mu$ used
previously by Reinders et al. \cite{Reinders+81,Reinders+85}. The scalar gluon
condensate 
$\langle\frac{\alpha_s}{\pi}G^2\rangle$ changes with density according to 
\begin{equation}
\langle\frac{\alpha_s}{\pi}G_{\mu\nu}G^{\mu\nu}\rangle_{n.m.}=\langle\frac{\alpha_s}{\pi}G_{\mu\nu}G^{\mu\nu}\rangle_{0}-\frac{8}{9}m_{N}^{0}\rho_N,
\end{equation}
where $m_{N}^{0}\simeq750\,\MeV$  is the nucleon mass in 
the chiral limit \cite{Borasoy+96}.

For the $J/\psi$ current, using  the background field technique \cite{Novikov+84}, we 
find that the additional contribution arising from the twist-2 operator looks
as follows,
\widetext
\Lrule
\begin{eqnarray}
\Delta\Pi_{\mu\nu}^V(q)& =&
\langle\frac{\alpha_s}{\pi} G^{ \alpha \sigma} G^\beta_{\,\,\,\sigma}
\rangle \frac{1}{Q^4} 
\biggl[ (-g_{\mu\nu}q_{\alpha }q_{\beta}+g_{\mu\alpha}q_{\nu}q_{\beta}+
q_{\mu}q_{\alpha}g_{\nu\beta}+g_{\mu\alpha}g_{\nu\beta}Q^2) \times \nonumber \\
&&  
\left(\frac{1}{2}+\left(1-\frac{Q^2}{3 m_c^2} \right)J_1-\frac{3}{2}J_2 \right) +   (g_{\mu\nu}-q_{\mu}q_{\nu}/q^2)q_\alpha q_\beta      
\left(-\frac{2}{3}+2J_1-2J_2+\frac{2}{3}J_3 \right) \biggl],
\end{eqnarray}
\Rrule
\narrowtext
\noindent
where $ J_{N}=\int_{0}^{1} dx [1+x(1-x)Q^2/m_c^2]^{-N}$.
In the present work we study the $c \bar{c}$-system at rest (relative to the
surrounding nuclear matter) and set $\vec{q}=0$, so that eqs.~(\ref{eq1}-\ref{eq2}) refer to
the Euclidean region $\omega^2=-Q^2<0$. Then there is only one invariant function,
\begin{equation}
\tilde{\Pi}^V(-Q^2=\omega^2)=-\frac{1}{3\omega^2} g^{\mu \nu} \Pi_{\mu
\nu}^V (\omega, \vec{q}=0), 
\end{equation}
which reduces to the usual vacuum polarization function when the  nuclear density goes to zero.

Similarly, for the pseudoscalar case, the gluonic twist-2 correction in the OPE
has the following form:  
\begin{eqnarray}
\Delta\Pi^P(q) &=& \langle\frac{\alpha_s}{\pi} G^{\alpha\sigma}
G^\beta_{\,\,\,\sigma} \rangle
\frac{q_{\alpha}q_{\beta}}{Q^4} \times \nonumber \\ && \left(\frac{1}{2}+\frac{1}{3}
 \left(1-\frac{Q^2}{m_c^2} \right) J_1-\frac{1}{6}
J_2-\frac{2}{3} J_3 \right)
\end{eqnarray}
Here we introduce the (dimensionless) polarization function
\begin{equation}
\tilde{\Pi}^P(-Q^2=\omega^2)=\frac{\Pi^P(\omega,\vec{q}=0)}{\omega^2}, 
\end{equation}
which reduces in the limit $\rho_N \to 0$ to the usual vacuum polarization function.

Our analysis is based on the moments of the polarization function
$\tilde{\Pi}^J$ with $J=V,P$ referring to the vector or pseudoscalar
channels. The $n$-th moment is connected, on the other side, with a dispersion
integral involving $\Im \tilde{\Pi}^J$,
\begin{eqnarray}
\label{dispersion}
M_n^J &\equiv& \left. { 1 \over n!} \left( {d \over d \omega^2} \right)^n
\tilde{\Pi}^{J}(\omega^2) \right|_{\omega^2=-Q_0^2} \nonumber \\ &&=
\frac{1}{\pi}\int_{4m_{c}^2}^{\infty}\frac{\mbox{Im}
\tilde{\Pi}^{J}(s)}{(s+Q_0^2)^{n+1}}ds,
\end{eqnarray}
at a fixed $Q_0^2=4m_c^2 \xi$. Direct evaluation of these moments using the
OPE gives
\begin{equation}
\label{opemoment}
M_n^J(\xi)= A^J_n(\xi) \left[ 1+ a^J_n(\xi) \alpha_s +b_n^J(\xi) \phi_b+c_n^J(\xi) \phi_c
\right].
\end{equation}
The common factor  $A^{J}_n$ results from the bare loop diagram. The coefficient $a_n^J$ takes into account perturbative radiative
corrections, while $b_n^J$ is associated with the gluon condensate term,
\begin{equation}
\phi_b =\frac{4 \pi^2}{9} { \langle \frac{\alpha_s}{\pi}G^2 \rangle
 \over (4m_c^2)^2 }, 
\end{equation}
The coefficients $A^{J}_n$,  $a_n^J$ and $b_n^J$ are listed in
ref.~\cite{Reinders+81}. The new contribution
from the twist-2 gluon operator involves
\begin{equation}
\phi_c  =  -\frac{2 \pi^2}{3} {\frac{\alpha_s}{\pi}
A_G \over (4m_c^2)^2 } m_N \rho_N.
\end{equation} 
For the additional Wilson coefficient $c_n$ we find in the vector channel:
\begin{equation}
  \label{cnv}
  c_n^V(\xi)=  b_n^V(\xi)-  \frac{4 n(n+1)}{3(2n+5) (1+\xi)^{2}} 
 { F(n+2,\frac{3}{2};n+\frac{7}{2};\frac{\xi}{1+\xi}) \over F(n,\frac{1}{2};n+\frac{5}{2}
;\frac{\xi}{1+\xi}) }    
\end{equation}
and in the pseudoscalar channel we obtain
\begin{equation}
  \label{cnp}
 c_n^P(\xi)=  b_n^P(\xi)-  \frac{4 n(n+1)}{  (1+\xi)} 
 { F(n+1,-\frac{1}{2};n+\frac{3}{2};\frac{\xi}{1+\xi}) \over F(n,\frac{1}{2};n+\frac{3}{2}
;\frac{\xi}{1+\xi}) } 
\end{equation}
with hypergeometric functions $F(a,b;c;z)$. By comparison with
ref.\cite{Reinders+81} we see that the $c_n$'s differ very little from the
$b_n$'s. From the resulting term $b_n^J(\phi_b+\phi_c)$ in
eq.~(\ref{opemoment}) one then observes that the gluon condensate effectively
changes by the following density dependent correction:
\begin{eqnarray}
  \label{approx}
  \langle \frac{\alpha_s}{\pi}G^2 \rangle_0 &\to& \langle \frac{\alpha_s}{\pi}G^2
  \rangle_0 - \left(\frac{8}{9} m_N^0+\frac{3}{2} m_N 
\frac{\alpha_s}{\pi}A_G \right) \rho_N
  \nonumber \\
&\simeq &\langle \frac{\alpha_s}{\pi}G^2 \rangle_0 (1-0.06 \rho_N / \rho_0),
\end{eqnarray}
using $\langle \frac{\alpha_s}{\pi}G^2 \rangle_0=(0.35  \GeV)^4$ and
$\rho_0=0.17 \, \fm^{-3}$.

The spectral function under the integral on the r.h.s. of eq.(\ref{dispersion})
is parameterized as 
\begin{equation}
\label{phen}
\Im \tilde{\Pi} (s)=\sum_i f_i
\delta(s-m_i^2) +  c \, \theta(s-s_0)\left(1+\frac{\alpha_s(s)}{4\pi} \right),
\end{equation}
in terms of a sum over low-lying resonances and a continuum part starting from
$s_0$, with $c=1/4 \pi^2$ in the vector channel and $c=3/8 \pi^2$ in the
pseudoscalar channel. In the vector channel the couplings $f_i$ of the $\bar{c}
\gamma_\mu c$ current to the $J/\psi,\,\psi',\,\psi'',\, ...$ resonances are
determined by their measured decay widths into $e^+ e^-$. Inserting eq.~(\ref{phen}) 
into eq.~(\ref{dispersion}), it is convenient to write 
\begin{equation}
\label{moments}
M_{n}^{J}(\xi)=\frac{f_0}{\pi (m^2+ Q_0^2)^{n+1} }
[1+\delta_{n}^{J}(\xi)],
\end{equation}
where $m$ is the mass of the lowest state, the one of interest. The
contributions of higher resonances as well as the continuum are absorbed in
$\delta_{n}^{J}$. Clearly, the relative importance of these higher energy parts
of the spectrum decreases with increasing $n$.
It is common practice to take the ratio of two neighboring moments,
$M_{n-1}/M_{n}= (m^2+Q_0^2)(1+\delta_{n-1})/(1+\delta_{n})$, so that $f_0$
drops out and one can focus on the mass $m$. For $n \geq 5$ it turns out that
$(1+\delta_{n-1}) /(1+\delta_{n})$ is close to one. 
Then the moment ratio does not depend on details of the higher resonances and
continuum parts of the spectrum, and we have
\begin{equation}
\label{ratio}
m^2 \simeq {M_{n-1}(\xi) \over  M_{n}(\xi) }-4 m_c^2 \xi.
\end{equation}

\noindent
\widetext
\begin{figure}[h]
\begin{picture}(500,200)
\put(0,0){\epsfig{file=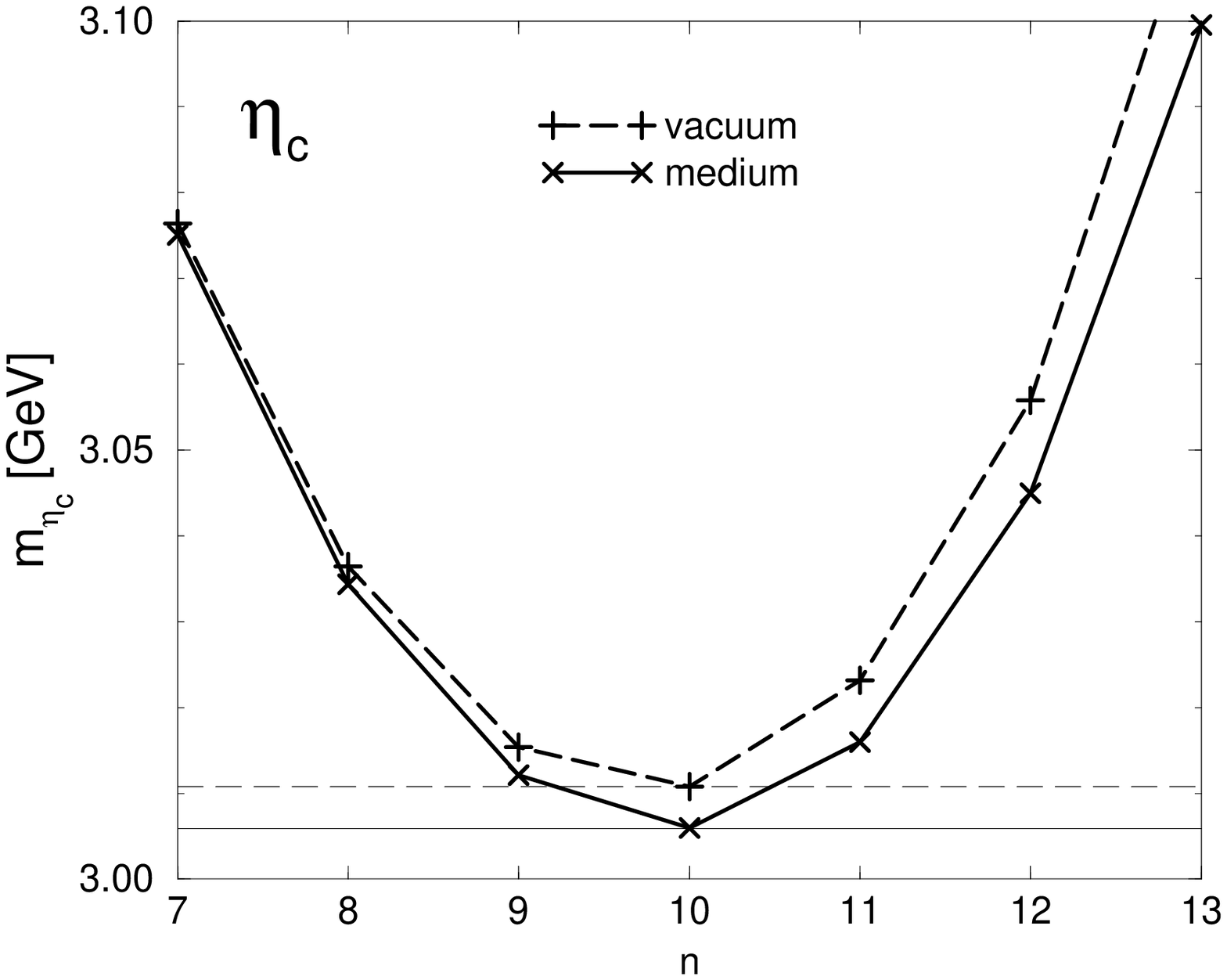,width=95mm}}
\put(245,0){\epsfig{file=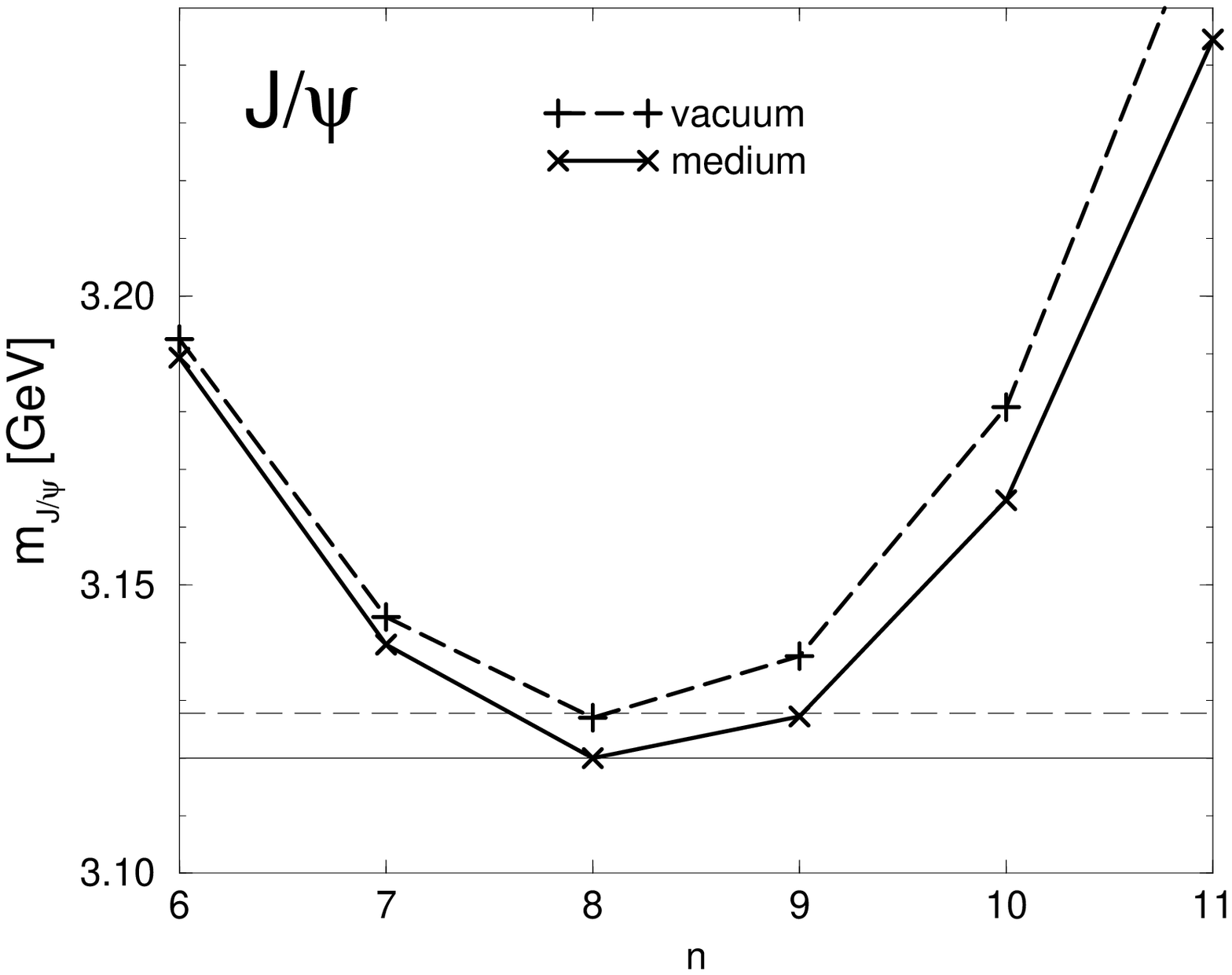,width=95mm}}
\end{picture}
\caption{The $\eta_c$ and $J/\psi$ masses calculated according to 
eq.~(\ref{ratio}) for
  different $n$ at $\xi=1$. We show the result in medium at $\rho_N=0.17
  \,\fm^{-3}$ (solid line) in comparison with the vacuum result (dashed
  line).}
\end{figure}
\narrowtext

The actual mass determination is done using moments in the range $7 \leq n\leq
11 $ and choosing $\xi=1$, just as in the vacuum case studied previously
\cite{Reinders+81,Reinders+85}. This range minimizes the sensitivity to details
of the high-energy spectrum. Going to larger $n$ would not
be justified without introducing additional, unknown condensates of higher
dimension in the OPE.

In Fig. 1 we show the results for in-medium masses (solid lines) of the
$J/\psi$ and $\eta_c$ at normal nuclear matter density $(\rho_N=\rho_0=0.17 \,
\fm^{-3})$ in comparison with their vacuum values (dashed lines). Using
$\alpha_s(8 m_c^2)=0.21$, $m_c=1.24\, \GeV$,
$\phi_b=1.8\cdot10^{-3}$ in the vacuum and $\phi_b=1.7\cdot10^{-3}$,
$\phi_c=-1.25\cdot10^{-5}$ in nuclear matter, we find the following mass shifts
taken at the minimal values of eq.~(\ref{ratio}):  
\begin{eqnarray}
  \Delta m_\psi \simeq -7 \, \MeV, \\
  \Delta m_{\eta_c} \simeq -5 \, \MeV. 
\end{eqnarray}
These shifts depend only very weakly on
our choice of parameters. Taking higher resonances explicitly into account
would lead to somewhat smaller mass shifts.

The sensitivity to possible enlarged in-medium widths of the $J/\psi$ or
$\eta_c$ turns out to be marginal. In fact no inelastic channels exist for
ground state $c\bar{c}$ mesons at rest interacting with a nucleon. Even if such
channels were present, they would not affect the mass shift analysis unless the
corresponding widths would reach magnitudes of 100 MeV or larger. This is in
qualitative contrast to light quark systems, such as the $\rho$ meson, for
which the in-medium width becomes so large that the QCD sum rule analysis of a
possible mass shift becomes ambiguous and inconclusive \cite{KKW97,LPM98}.

In summary our in-medium QCD sum rule analysis, with the operator product
expansion calculated up to dimension four, predicts attractive mass shifts of
about 5-10 MeV for $J/\psi$ and $\eta_c$ in nuclear matter. This corresponds
to small $J/\psi$- and $\eta_c$-nucleon scattering lengths $a=-\mu_r \Delta m/(2 \pi
\rho_N) \simeq (0.1-0.2)\, \fm$ ($\mu_r$ is the meson-nucleon reduced mass). 
Our results for the mass shifts of the lowest $\bar{c}c$ states are
surprisingly close to those reported in ref. \cite{Luke+92,Brodsky+97,Teramond+98}. 
Most of the calculated mass shift comes from the 
density dependence of the gluon condensate.  
The new term related to the fraction of momentum
carried by gluons in the nucleon contributes  less than 10\% to the total 
effect.  The influence of
the decay widths is expected to be very small, at least for $J/\psi$ and
$\eta_c$ at rest\cite{Brodsky+97}. 
Of course, for charmonium systems traversing nuclear matter 
at high energy, the scattering amplitudes can develop substantial imaginary 
parts from reactions with nucleons producing open charm \cite{Kharzeev94}.

After submission of this paper a similar calculation has been reported in ref. \cite{Hayashigaki98}

\widetext

\begin{thebibliography}{99}
\bibliographystyle{unsrt}

\setlength{\itemsep}{0.0in}

\bibitem{Brodsky+90} S.J. Brodsky, I. Schmidt and G.F. de Teramond,
Phys.\,Rev.\,Lett.\  {\bf 64}, 1011 (1990).

\bibitem{Wasson91} D.A. Wasson,
Phys.\,Rev.\,Lett.\  {\bf 67}, 2237 (1991). 

\bibitem{Luke+92} M. Luke, A.V. Manohar, M.J. Savage,
Phys.\,Lett.\  B {\bf 288}, 355 (1992).

\bibitem{Brodsky+97} S.J. Brodsky, G.A. Miller,
Phys.\,Lett.\ B {\bf 412}, 125 (1997).

\bibitem{Teramond+98} G.F. de Teramond, R. Espinoza, M. Ortega-Rodriguez,
Phys.\,Rev.\ D {\bf 58}, 034012 (1998).

\bibitem{Shifman+79} M.A. Shifman, A.I. Vainshtein and V.I. Zakharov, 
Nucl.\,Phys.\, {\bf B147}, 385 (1979); Nucl.\,Phys.\, {\bf B147}, 448 (1979).

\bibitem{DL91} E.G.~Drukarev, E.M.~Levin, Prog.~Part.~Nucl.~Phys.~{\bf 27}, 77 
(1991). 

\bibitem{Hat91} T.~Hatsuda, H.~Hogaasen, M.~Prakash, 
Phys. Rev. Lett. {\bf 66}, 2851 (1991). 

\bibitem{Tom91} T.D.~Cohen, R.J.~Furnstahl, D.K.~Griegel, 
Phys. Rev. Lett. {\bf 67}, 961 (1991). 

\bibitem{HL92} T.~Hatsuda and S.H.~Lee,
Phys.\,Rev.\, C {\bf 46}, R34 (1992).

\bibitem{Wilson69} K.G. Wilson, Phys. Rev. {\bf 179}, 1499 (1969).

\bibitem{Tom95} T.D.~Cohen, R.J.~Furnstahl, D.K.~Griegel and  X.-M. Jin,  
Prog.~Part.~Nucl.~Phys.~{\bf 35}, 221 (1995).  

\bibitem{Reinders+81} L.J. Reinders, H.R. Rubinstein and S. Yazaki,
Nucl.~Phys.~{\bf B186}, 109 (1981).

\bibitem{Reinders+85} L.J. Reinders, H.R. Rubinstein and S. Yazaki,
Phys.~Rep.~{\bf 127}, 1 (1985).

\bibitem{Cohen+92} T.D. Cohen, R.J. Furnstahl, D.K. Griegel, 
Phys.\,Rev.\ C {\bf 45}, 1881 (1992).  

\bibitem{Gluck+} M. Gl\"uck, E. Reya, A. Vogt, 
Z.\,Phys.\ C {\bf 53}, 127 (1992). 

\bibitem{Borasoy+96} B. Borasoy and U.-G. Meissner, 
Phys.\,Lett.\ {\bf 365}, 285 (1996).

\bibitem{Novikov+84} V.A. Novikov, M.A. Shifman, A.I. Vainshtein and 
 V.I. Zakharov, Fortschr.\,Phys.\ {\bf 32}, 585 (1984); A.~V.~Smilga,
 Sov.~J.~Nucl.~Phys.~{\bf 35}, 271 (1982).

\bibitem{KKW97}    {F. Klingl, N. Kaiser and W. Weise,
                     Nucl. Phys. {\bf A624}, 527 (1997).}

\bibitem{LPM98}     {S. Leupold, W. Peters and U. Mosel,
    Nucl.\ Phys.\ {\bf A628}, 311 (1998).}

\bibitem{Kharzeev94} D.~Kharzeev and H.~Satz, Phys.~Lett.~B {\bf 334} (1994) 155. 

\bibitem{Hayashigaki98} A. Hayashigaki, 
nucl-th/9811092.
 
\end{thebibliography}
\end{document}